\documentclass[prb,floatfix,twocolumn]{revtex4}
\usepackage[dvips]{color}
\usepackage{graphicx}
\usepackage{rotating,amssymb}
\begin{document}
\title{Jahn-Teller like origin of the tetragonal distortion\\ 
in disordered Fe-Pd magnetic shape memory alloys}
\author{Ingo Opahle$^{1,2}$, Klaus Koepernik$^1$, Ulrike Nitzsche$^1$ and Manuel Richter$^1$}
\affiliation{$^1$IFW Dresden, P.O.B. 270016, D-01171 Dresden, Germany\\
$^2$Institut f\"ur Theoretische Physik, Universit\"at Frankfurt, 60438 Frankfurt/Main, Germany}
\date{\today}
\begin{abstract}
The electronic structure and magnetic properties of disordered 
Fe$_{x}$Pd$_{100-x}$ alloys $(50 < x < 85)$ are investigated in the 
framework of density functional theory using the full potential 
local orbital method (FPLO). 
Disorder is treated in the
coherent potential approximation (CPA). Our calculations 
explain the experimental magnetization data.
The origin of the tetragonal distortion in the Fe-Pd magnetic
shape memory alloys 
is found to be a Jahn-Teller like
effect which allows the system to reduce its band energy in a narrow 
composition range.
Prospects for an optimization of
the alloys' properties by adding third elements are discussed.
\end{abstract}
\maketitle
Magnetic shape memory (MSM) alloys have attracted considerable attention as
materials for actuator and sensor applications, due to large magnetically
induced strains of up to 9.5\% in Ni$_2$MnGa.~\cite{Soz02}
A promising MSM alloy is disordered Fe$_{70}$Pd$_{30}$~\cite{Jam98}
with a relatively high blocking stress, a high saturation magnetization 
and a Curie temperature well above room temperature, 
as well as a high ductility, in contrast to Ni$_2$MnGa. To date,
magnetic field induced strains of about 3\% have been observed in 
Fe-Pd alloys.~\cite{Kak06a} 

The unusually large strains observed in Ni$_2$MnGa and Fe$_{70}$Pd$_{30}$,
which are one to two orders of magnitude higher than in conventional
magnetostrictive materials, are attributed to a reorientation of 
martensitic twins in the magnetic field. Twin variants which have their 
easy axis of magnetization aligned along the field direction gain
Zeeman energy and grow at the cost of variants with an unfavorable 
alignment. This effect thus requires 
a magnetocrystalline anisotropy energy large enough to prevent 
a rotation of the magnetization within a twin variant and a 
high mobility of twin boundaries.

In this Letter, we investigate the electronic structure of
disordered Fe$_{x}$Pd$_{100-x}$ alloys $(50 < x < 85)$ by means of
density functional theory calculations. In particular, we explore
the electronic origin of the tetragonal distortion resulting in the fct martensite
phase, where the MSM effect is observed.
On the basis of these results, we discuss prospects for a further
optimization of the alloys' properties.

The value of the $c/a$-ratio in the fct martensite phase
determines the MSM properties of Fe-Pd alloys in a twofold way:
First, the maximum strain is given by $\epsilon = \Delta l/l = |1-c/a|$
under the assumption of a hundred percent conversion of twin variants.
Second, the magnetocrystalline anisotropy energy, which determines the 
blocking stress, is expected to depend strongly on the 
$c/a$-ratio.~\cite{Lyu05,Bus07} Thus, understanding the mechanism behind
the tetragonal distortion is important for the development of alloys
with improved MSM properties.

The martensitic transformation from fcc to fct in disordered Fe-Pd alloys 
is observed in a narrow composition range between 
29 and 32 at\% Pd,~\cite{Osh82,Sug84}
close to the transition from an fcc to a bcc 
ground state with increasing Fe content.
Upon further cooling an irreversible transition from fct to a (bcc-like)
bct phase occurs.
The martensite temperature $T_m$ is 
close to room temperature for optimal alloy composition, but depends 
strongly on the Fe content.~\cite{Sug84}
For applications a martensite temperature $T_m$ clearly
above room temperature would be desirable, since the operation range
of MSM elements is limited by $T_m$. 
Attempts to raise $T_m$ in the Fe-Pd system by addition of Co, Ni, Rh or
Pt have so far been unsuccessful.~\cite{Tsu03,Vok05,Wad03,Osh82}
Furthermore, the $c/a$-ratio
in the martensite phase depends both on composition and on temperature.
A gradual increase of $c/a$ from about 0.92 at -80$^o$ C to 0.97
near $T_m$ was observed,~\cite{Osh82} similar to the temperature
dependence of $c/a$ in Fe$_3$Pt, 
but distinct from the prototype MSM alloy Ni$_2$MnGa.~\cite{Kak06a}

We have calculated the electronic structure of disordered 
Fe$_{x}$Pd$_{100-x}$ alloys ($50<x<85$) in the framework of density
functional theory (DFT) in the local spin density approximation (LSDA)
using the high precision full potential local orbital (FPLO, version 5.00-19) 
method~\cite{FPLO} in the scalar relativistic approximation. 
Disorder was treated within the coherent potential
approximation (CPA)~\cite{CPA} and the LSDA parameterization 
of Perdew and Wang~\cite{Per92}
was employed. The basis set consisted of Fe 3spd4sp and Pd 4spd5sp
optimized local orbitals. The exponent of the confining potential
in the basis optimization was set to $n=6$, which provided lower
total energies and a better numerical stability compared to the default
value $n=4$.
For the {\bf k}-space integrations 1728 points in the full
Brillouin zone (FBZ) were used, and the results were cross-checked with 8000
points in the FBZ. Additionally, we have performed calculations
with the Akai-KKR code~\cite{AkaiKKR} in the muffin tin approximation
using the generalized gradient approximation (GGA)
to DFT in the parameterization by Perdew {\em et al.}~\cite{Per92a}

The calculated magnetic moments of fcc Fe$_{x}$Pd$_{100-x}$ alloys
$(50<x<70)$ are in an overall good agreement with the experimental 
data by Matsui {\em et al.}~\cite{Mat89}
The average spin moment per atom, calculated with FPLO at the respective
experimental lattice parameter, increases approximately linear with increasing 
Fe content from 1.55 $\mu_B$ for $x=50$ to 1.98 $\mu_B$ for $x=70$.
Adding the contributions of the orbital moments, which are estimated 
from relativistic supercell calculations~\cite{Bus07} to be about 0.13 $\mu_B$ 
for Fe and 0.04 $\mu_B$ for Pd, the calculated moments
agree within 0.05 $\mu_B$ with the experimental values for the average total 
moments 1.61 $\mu_B$ ($x=50$) and 2.04 $\mu_B$ ($x=70$).

Assuming that the individual atomic moments do not depend on the
compostion, Matsui {\em et al.}~\cite{Mat89} had to use an unusually large
Pd moment of 0.56 $\mu_B$ to fit the concentration dependence
of their magnetization data. This value is nearly twice the value we obtained for the
Pd spin moment (0.29 $\mu_B$).
Our calculations provide a different explanation for the observed
concentration dependence.
The spin moment of Fe decreases with increasing Fe
content due to magneto volume effects, from about 2.82 $\mu_B$ for 
$x=50$ to 2.71 $\mu_B$ for $x=70$.
A yet more pronounced reduction of the Fe moments with increasing
Fe content was recently observed in neutron scattering experiments
on L1$_0$-ordered FePt alloys~\cite{Lyu06}.

\begin{figure}
\includegraphics[width=0.45\textwidth]{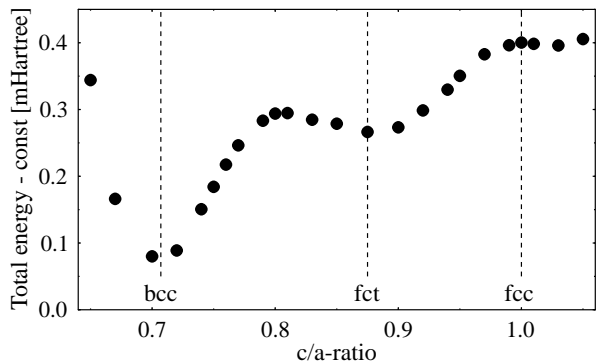}
\caption{\label{FIG:bain}Total energy as a function of $c/a$-ratio
for Fe$_{80}$Pd$_{20}$ calculated with the atomic volume fixed to
$V_{\rm at}$=13.256 \AA$^3$. The total energy exhibits a local minimum
at $c/a$=0.875 corresponding to an fct martensite structure.}
\end{figure}

The fcc structure can be continuously transformed into a bcc structure 
by a change of the $c/a$-ratio, the so called Bain path. A (meta) stable 
fct phase must exhibit a (local) minimum in the free energy along the 
Bain path for $0.71 < c/a <1$. In our total energy
calculations we find such a local minimum only in a narrow region of 
phase space, where the total energy differences between fcc and bcc 
are small and the energy landscape along the Bain path is 
accordingly flat~(Fig.~\ref{FIG:bain}). This is in agreement with
the experimentally observed proximity of the martensite phase to
the transition from an fcc to a bcc ground state.
While in the LSDA calculations the fcc-bcc transition is found at a somewhat higher Fe 
concentration $x \approx 80$  in the vicinity of the experimental volume,
the KKR GGA data yield almost exactly the experimental value $x \approx 70$.

\begin{figure}
\includegraphics[width=0.43\textwidth]{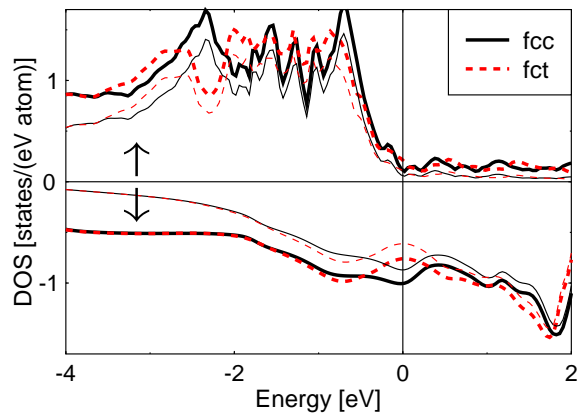}
\caption{\label{FIG:DOS}(Color online) 
DOS of Fe$_{80}$Pd$_{20}$ in the
fcc (solid lines) and fct (dashed lines) structures corresponding to
Fig.~\ref{FIG:bain}. The total DOS (thick lines) and the partial Fe 3d
DOS (thin lines) are shown. The minority spin DOS~($\downarrow$)
exhibits a peak at the Fermi energy (vertical line) in fcc which splits
in fct.}
\end{figure}
Further insight into the origin of the martensite phase
is obtained from the density of states (DOS) shown in Fig.~\ref{FIG:DOS}.
The DOS in the vicinity of the Fermi energy is dominated by Fe 3d states,
which are hybridized with Pd 4d states (not shown). 
The majority spin channel of the Fe 3d subshell is completely filled,
while in the partially filled minority spin channel a peak is located 
right at the Fermi energy in the fcc austenite phase.
In the fct martensite phase this peak is split and weight is shifted 
below and above the Fermi energy, indicative of a cooperative Jahn-Teller 
effect. The gain in band energy due to this splitting
(estimated from a single step calculation) is a few tenths of a mHartree.

\begin{figure}
\includegraphics[width=0.45\textwidth]{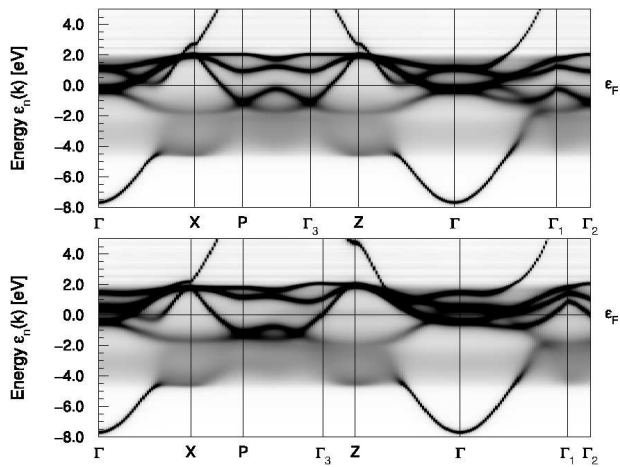}
\caption{\label{FIG:bands}CPA minority spin spectral densities 
of Fe$_{80}$Pd$_{20}$ in the fcc structure (upper panel) and fct
(lower panel) corresponding to Fig.~\ref{FIG:bain}.}
\end{figure}

Further support for a Jahn-Teller like origin of the tetragonal distortion
in the disordered Fe-Pd MSM alloys is obtained from the spectral
density of the minority spin states shown in Fig.~\ref{FIG:bands}.
For the cubic austenite phase (upper panel) the threefold
degenerate $t_{2g}$ states of the Fe 3d minority spin shell are located
right at the Fermi level close to the $\Gamma$-point (note that due to 
disorder a broadening of the bands occurs).
The tetragonal distortion in the martensite phase (lower panel) causes
a splitting of the $t_{2g}$ states, which are shifted away from the Fermi
energy, resulting in a gain in band energy in a Jahn-Teller like fashion.

On the basis of the results discussed above, the following conclusions
can be drawn: The presence of degenerate Fe 3d minority spin bands
at the Fermi energy in Fe-Pd MSM alloys favors a structural distortion, 
which is stabilized by a gain in band energy. However, due to band dispersion
and disorder, this energy gain is small and thus can only
compete with other energy terms (like the Madelung energy favoring close
packed structures) if the energy landscape is sufficiently flat,
i.e. close to the fcc-bcc transition. 
The gain in band energy due to the Jahn-Teller like splitting
is approximately proportional to the height of the peak in the minority
spin DOS. The energy difference between fcc and bcc changes by approximately 4 meV/at\%, 
while the height of the peak in the DOS shows only a moderate concentration
dependence.

Thus already small deviations from the optimal stoichometry 
result in relatively steep energy curves along the Bain path and cause the local
fct minimum to disappear. 
Consequently, a stabilization 
of the fct structure in terms of $c/a$-ratio and martensite temperature $T_m$
is expected with increasing iron concentration $x$, up to the point where
a bcc (or bct) structure becomes favorable. Alternatively, alloying Fe-Pd with
ternary elements could lead to a stabilization of the fct phase.
According to our KKR GGA calculations, Fe-Pd-X alloys with about 5 at\% of the ternary elements X= W, Au or Pt 
show a more pronounced peak in the minority spin DOS close to the respective
fcc-bcc transition and are therefore
interesting candidates for MSM alloys with enhanced $T_m$.

\begin{figure}[t!]
\includegraphics[width=0.45\textwidth]{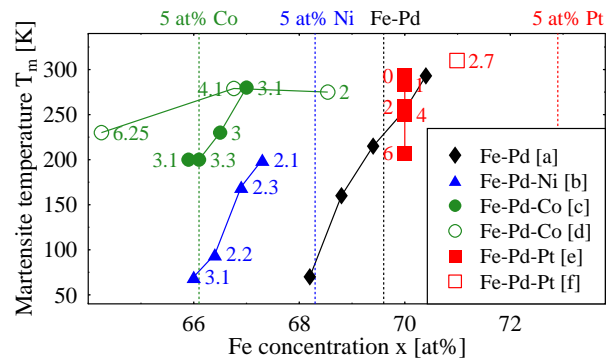}
\caption{\label{FIG:fepdx}Experimental martensite temperature as a function 
of the Fe concentration $x$ for Fe$_x$Pd$_{100-x}$ 
(a:~Ref.~\onlinecite{Sug84}),
Fe$_x$Pd$_{100-x-y}$Ni$_y$ (b:~Ref.~\onlinecite{Tsu03}),
Fe$_x$Pd$_{100-x-y}$Co$_y$ (c:~Ref.~\onlinecite{Tsu03}; 
d:~Ref.~\onlinecite{Vok05}),
and Fe$_x$Pd$_{100-x-y}$Pt$_y$ (e:~Ref.~\onlinecite{Wad03},
f:~Ref.~\onlinecite{Vok05}).
Numbers near data points indicate the amount of ternary element $y$.
Also shown is the calculated Fe concentration $x_c$ for the fcc-bcc 
transition (dashed vertical lines) for Fe-Pd and ternary alloys with
$y$= 5 at\%.
}
\end{figure}

Our results imply that $T_m$ and $c/a$ 
in Fe-Pd(-X) alloys are determined by a delicate balance 
between the energy gain due to the Jahn-Teller splitting
and the proximity to the fcc-bcc transition. 
This is in line with the
experimentally observed $T_m$ of binary and ternary Fe-Pd(-X) 
MSM alloys (X= Co, Ni, Pt), which shows a strong increase with
the Fe concentration $x$ for alloys having a comparable amount of the ternary element 
X (Fig.~\ref{FIG:fepdx}).
The critical Fe concentration $x_{\rm c}$, 
where fcc and bcc are degenerate in energy
depends on the amount and the type of X (Fig.~\ref{FIG:fepdx}).
The values of
$x_{\rm c}$ have been obtained from KKR GGA calculations and depend below
10 at\% of X approximately linear on the amount of ternary element.
The KKR GGA calculations yield $x_{\rm c}$ for
the binary alloys in excellent agreement with the experimental trends, and the
relative trends upon alloying can be considered trustworthy despite
the limited accuracy of the muffin tin approximation.
For Co and Ni, $x_{\rm c}$ is shifted to lower values, while for Pt it 
is increased with respect to the binary system. This naturally explains
that in the experiments of Ref.~\onlinecite{Wad03} addition of Pt at constant
Fe content was found to have a negative influence on $T_m$, as with increasing
Pt content $x_{\rm c}$ was shifted away from 70 at\% Fe concentration used.
The somewhat higher value of $T_m$ observed in Ref.~\onlinecite{Vok05} is in line with our argument.
Consequently, a careful optimization of the Fe content is necessary to achieve $T_m$ enhancement in
ternary Fe-Pd-X alloys.

This work was supported financially by the DFG, SPP 1239.
Discussions with Monodeep Chakraborty are gratefully acknowledged.

\end{document}